\documentclass[12pt,preprint]{aastex}
\def\({\left(}
\def\){\right)}
\def\[{\left[}
\def\]{\right]}
\def\msun{{\rm ~M}_{\odot}}

\def\ledd{{\rm L_{Edd}}}
\usepackage{graphicx}
\usepackage{natbib}
\usepackage[usenames,dvips]{color}

\shorttitle {Inner Cool Disks in the Low Hard State}
\begin{document}

\title{The Existence of Inner Cool Disks in the Low Hard State of 
Accreting Black Holes}

\author{B. F. Liu}
\affil{National Astronomical Observatories/Yunnan Observatory, Chinese 
Academy of Sciences, P.O. Box 110, Kunming 650011, China}
\email{bfliu@ynao.ac.cn}

\author{Ronald E. Taam}
\affil{Northwestern University, Department of Physics and Astronomy,
  2131 Tech Drive, Evanston, IL 60208; ASIAA/National Tsing Hua University -
 TIARA, Hsinchu, Taiwan}
\email{r-taam@northwestern.edu}

\author{E. Meyer-Hofmeister} 
\affil{Max-Planck-Institut f\"ur Astrophysik, Karl Schwarzschildstr. 1, 
D-85740, Garching, Germany}
\email{emm@mpa-garching.mpg.de}

\and 

\author{F. Meyer} 
\affil{Max-Planck-Institut f\"ur Astrophysik, Karl Schwarzschildstr. 1, 
D-85740, Garching, Germany}
\email{frm@mpa-garching.mpg.de}

\begin{abstract}
The condensation of matter from a corona to a cool, optically thick 
inner disk is investigated for black hole X-ray transient systems in 
the low hard state.  A description of a simple model for the exchange 
of energy and mass between corona and disk originating from  thermal 
conduction is presented, taking into account the effect of Compton 
cooling of the corona by photons from the underlying disk. It is found 
that a weak, condensation-fed inner disk can be present in the low hard 
state of black hole transient systems for a range of luminosities 
which depend on the magnitude of the viscosity parameter.  For $\alpha 
\sim 0.1-0.4$ an inner disk can exist for luminosities in the range $\sim 
0.001- 0.02 \ledd$.  The model is applied to the X-ray observations of the 
black hole candidate sources GX 339-4 and Swift J1753.5-0127 in 
their low hard state.  It is found that Compton 
cooling is important  in the condensation process, leading 
to the maintenance of cool inner disks in both systems. 
As the results of the evaporation/condensation model are independent of 
the black hole mass, it is suggested that such inner cool disks may 
contribute to the optical and ultraviolet emission of low luminosity active 
galactic nuclei. 
\end{abstract}


\section{Introduction}

Black hole X-ray transient binary systems have attracted increasing attention 
in recent years since they can be used as a probe of the underlying physics of the 
accretion process in disks surrounding black holes over a wide range in luminosity.  
The X-ray spectral behavior of these systems is complex, exhibiting differing states 
and transitions.  In particular, it is well known that two basic X-ray spectral 
states are present with a soft spectral state occurring at high luminosities and a 
hard spectral state occurring at low luminosities. The properties of these 
systems have been reviewed by Tanaka \& Shibazaki (1996) and more recently 
by Remillard \& McClintock (2006) and McClintock \& Remillard 
(2006).  It is now generally accepted that these two spectral states originate from 
different accretion modes dependent on the mass accretion rate.

At high luminosities, black hole X-ray transients are characterized 
by a soft thermal spectrum described by a multi-color black body component 
dominant at about 1 keV.  This has been interpreted as arising from an optically 
thick accretion disk extending to the innermost stable circular orbit (Shakura \& 
Sunyaev 1973).  In contrast, at low luminosities, the systems are characterized 
by a hard spectral state where the spectrum is described by a power law with 
a typical photon index of about 1.7. The emission is commonly 
thought to be produced by the Compton scattering of soft photons with 
the hot electrons in an optically thin inner disk (Shapiro et al. 
1976; Sunyaev \& Titarchuk 1980; Pozdnyaknov et al. 1983).  In these models, 
the ions and electrons are described by a two temperature plasma.  Here, the ions 
are heated to high temperatures by viscous dissipation, the electrons to attain  lower 
temperatures due to their strong interaction with radiation and weak Coulomb coupling 
with ions.  The initial models developed for this 
state (Shapiro et al. 1976) were thermally unstable, and it has been 
recognized that the radial advection of internal energy can stabilize the 
flow (see, for example, Narayan 2005). In this case, the internal energy is advected 
inward with the flow, resulting in an inefficient conversion of 
gravitational potential energy to radiation (e.g., Narayan \& Yi 1994; Narayan 
\& Yi 1995a,b).  The radius of transition between the cool outer disk and the 
hot advection-dominated inner disk was not theoretically determined, but obtained 
by fits to the observational data (see Esin et al. 1997).  

\begin{figure}
\plotone{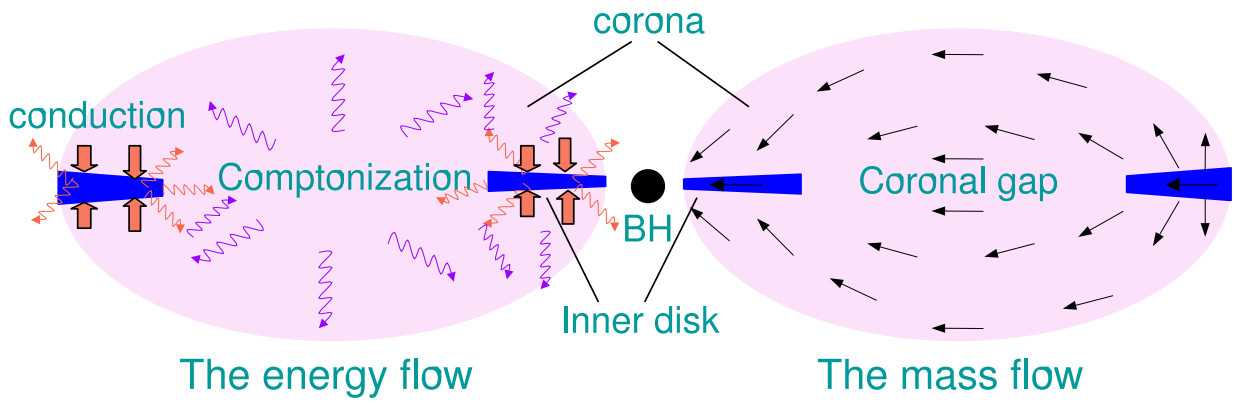}
\caption{Schematic picture of the truncated outer disk separated from the inner
disk by a coronal gap, indicating the energy and mass flow in the configuration.}
\label{f:dsk-cor.ps}
\end{figure}

The transition between these states occurs at luminosities in the range of 
$\sim 1-4\%$ of the Eddington value (Maccarone 2003) and is thought to be a 
consequence of a disk corona interaction (Meyer, Liu, \& Meyer-Hofmeister 2000b). 
More detailed observations and analyses indicated that, in addition to the soft 
and hard state, an intermediate state could occur during the rise or 
decay of an outburst between the hard and soft state.  Here, both a soft 
thermal spectrum and a hard power law spectrum are observed, likely indicating 
the coexistence of a hot optically thin and cool optically thick disk structure.  
Such a disk structure can be envisaged if the optically thick disk is truncated 
by some evaporative process, leading to the formation of an inner hot geometrically 
thick, optically thin disk surrounded by an outer cool geometrically thin disk. Originally, 
the idea of a two phase accretion structure was suggested by Eardley et al. 
(1975) and Shapiro et al. (1976).  Such disk geometries have recently been 
calculated based on a proton bombardment model (Dullemond \& Spruit 2005) 
and a coronal evaporation model controlled by electron conduction (Liu et al. 
1999, R\`o\.za\`nska \& Czerny 2000a,b). In the latter case a maximum evaporation
rate was found which allowed an estimate of a smallest transition radius 
and a maximal mass flow rate for which the evaporation rate into the corona can still exceed the mass flow 
rate in the cool outer disk (Liu et al. 1999). Alternatively Liu, Meyer, 
\& Meyer-Hofmeister (2006) and  Meyer, Liu, \& Meyer-Hofmeister (2007)  
have suggested that a remnant cool inner disk formed during the thermally dominant (soft) spectral 
state, when the accretion rate declines just below the transition rate, can be maintained for times longer than a viscous diffusion time by
condensation of matter from the corona.  The accretion geometry then would be described as a cool inner and even cooler outer 
disk separated by a gap filled with an advection dominated accretion flow (ADAF) 
(Fig. \ref{f:dsk-cor.ps}, see also Mayer \& Pringle 2007). Here, the inner cool 
disk is responsible for the presence of the soft thermal 
component coexisting with the hard component formed in the coronal region.  

Recently, evidence pointing to the possible presence of a cool 
inner disk during the low hard state of black hole transient systems has been 
provided by observations of GX 339-4 and Swift J1753.5-0127 (Miller et al. 
2006a,b).  In particular, Miller et al. (2006a) find that a soft thermal 
component ($kT \sim 0.2$ keV) is required to fit the spectrum of J1753.5-0127, 
with a normalization suggesting a small inner disk region. Similarly, a soft 
thermal component characterized by kT $\sim 0.3$ keV was required for GX 339-4
(Miller et al. 2006b).  In this latter case, a broad Fe K line was also required to 
fit the spectrum, providing additional evidence to support the hypothesis of cold matter 
lying close to the black hole.  We note that this interpretation is model dependent 
since it is possible that the emission line may have formed in an outflowing wind 
(Laurent \& Titarchuk 2007).  Nevertheless, the existence of such an inner cold disk 
region is very suggestive.  Such a picture may be consistent with the accretion geometry 
envisaged by Liu et al. (2006) and Meyer et al. (2007), or cold clumps in luminous hot accretion flows (Yuan 2003). However, Miller et al. 
(2006a) find evidence for a soft component from a weak disk extending to the 
innermost stable circular orbit (ISCO) at X-ray luminosities as low as 0.003 
$\ledd$, significantly less than the luminosity corresponding to the transition 
between the high soft and low hard states.  

In this paper, we investigate the properties of a cool inner disk within the framework of 
the coronal evaporation and recondensation model, extending earlier work 
by Liu et al. (2006) and Meyer et al. (2007), to determine whether cool 
disks are present at low X-ray luminosities.  In contrast to Liu et al. 
(2006) and Meyer et al. (2007), we include the cooling effect associated 
with the inverse Compton scattering of photons emitted by an underlying 
disk on hot thermal electrons for determining the thermal state of the 
corona.  Assuming that thermal conduction determines the energy and mass 
exchange between corona and underlying cool disk, we examine the range of 
conditions under which a cool inner disk forms and is maintained.  In the 
next section we give an overview of the evaporation/condensation model, 
including the additional input physics.  In \S 3 illustrative results are 
presented for the case where the cooling of electrons is dominated by heat 
conduction to the underlying disk and for the case where electron cooling 
is dominated by Compton scattering of cool photons. To illustrate the 
potential applicability of these concepts to observed sources, we compare 
the numerical results to the X-ray observations of the black hole candidate 
source GX 339-4 and the recently discovered system Swift J1753.5-0127 
(Palmer et al. 2005) in \S 4.  Finally, we discuss our results and their 
implications in the last section.

\section{Theoretical Model}

The formation of a corona above a geometrically thin disk may result from 
physical processes similar to those operating in the solar corona or from 
a thermal instability (e.g., Shaviv \& Wehrse 1986) in the uppermost 
layers of a disk.  As mentioned above, earlier work 
(Meyer \& Meyer-Hofmeister 1994; Meyer, Liu, \& Meyer-Hofmeister 2000a, Liu 
et al. 2002; R\`o\.za\'nska \& Czerny 2000a,b) revealed that a disk corona
fed by the evaporation of matter from an underlying cool disk can be maintained
as an accreting coronal flow.  The description of the mechanism responsible for 
the evaporation is briefly described below.  

In the corona, the viscous dissipation leads to ion heating, which is partially 
transferred to the electrons via Coulomb collisions.  This energy is assumed 
to be conducted into a lower, cooler transition layer. If the 
density in this layer is sufficiently high, the conductive flux is radiated away.  
On the other hand, if the density is insufficient to efficiently radiate the energy, 
the underlying cool gas is heated and evaporation into the corona takes place.
The evaporated matter carries angular momentum, and it gradually accretes onto 
the central black hole as a result of viscous transport.  As the accreted gas 
is resupplied by the material evaporated from the cool disk, a steady accreting 
corona is formed above the disk in which a balance of evaporation and accretion 
is achieved. 

While the disk evaporation provides matter for accretion in the corona, the mass 
accretion through the cool outer disk is decreased and can even vanish at distances 
where the inflowing rate is lower than the evaporation rate, 
approximated by $\dot M_{\rm evap}
\equiv 4\pi R^2\dot m_0$, where $\dot m_0$ is the evaporation rate per unit area 
from the disk into the corona.  Numerical calculations (Meyer et al. 2000a; Liu et 
al. 2002) reveal that the  evaporation rate increases with decreasing distance 
to the central object until a maximum evaporation rate of about a percent of the 
Eddington rate is reached at a distance of several hundred Schwarzschild radii.
The existence of such a maximum leads to a change in character of the accretion 
geometry. If the mass flow in the disk is below this maximum value, as in the 
quiescent state of black hole X-ray transients, the optically thick disk is truncated 
at the distance where all matter is evaporated, leaving a pure advection dominated
coronal flow.  As the mass flow rate increases (e.g., during the rise to an 
outburst), the edge of the cool optically thick disk moves inward. If the mass 
flow in the disk increases above the maximal evaporation rate, the disk can not be 
evaporated completely and, hence, the cool disk extends to the ISCO.  During the 
decline from the peak of the outburst, the inner edge of the cool disk retreats 
outwards to greater distances from the central black hole. These variations in 
the accretion process during the outburst of a black hole X-ray transient 
system provide explanation for the hard and soft states in these systems (Meyer 
et al. 2000b).  Recent investigations (Liu et al. 2006; Meyer et al. 2007) 
furthermore reveal that when the accretion rate is not far below the maximal 
evaporation rate an inner disk separated from the outer disk by a coronal region 
could also exist, leading to an intermediate state of black hole accretion.  

The onset of an intermediate state occurs as the accretion rate decreases just 
below the maximal evaporation rate. At this time, disk truncation by evaporation 
sets in near the region where the evaporation rate is maximal. A coronal gap 
appears and widens with a further decrease in the accretion rate with the inner 
cool disk reduced in extent. Because of diffusion, the inner disk can not survive 
for a time longer than a viscous time (which is only a few days in the inner 
disk) unless matter continuously condenses from the ADAF onto the cool inner disk. 
In the following we investigate the interaction between the disk and the corona/ADAF, 
showing the conditions under which matter condenses onto the inner disk, thereby 
maintaining a cool disk in the inner region.                                                                                    
\subsection{Model Assumptions}

The corona lying above an inner cool disk is similar to an ADAF. The main physical
differences between these two descriptions stem from the existence of vertical 
thermal conduction caused by the large temperature gradient between the corona 
and the disk and the presence of cool disk photons, which upon propagating through 
the corona remove energy by inverse Compton scattering. At low accretion 
rates, neither conduction nor Compton scattering is important for energy loss and the corona is 
described by an ADAF. In this study, we assume that the corona above the cool 
disk can be described by an ADAF, where the structure (such as pressure, density, 
and ion temperature) is determined by the mass of the black hole, the mass 
accretion rate, and the viscosity at a given distance (Narayan \& Yi 1995a).  
However, the cooling in such an ADAF is assumed to be either dominated by the 
Compton scattering of the disk photons or dominated by vertical heat conduction. That 
is, we implicitly assume that non-thermal (e.g., synchrotron) processes 
are unimportant in the lowest 
level of approximation. As we shall see, such a simple model contains the 
ingredients necessary to provide an understanding of the soft spectrum in the 
low/hard state.  The model allows an analytical description, revealing the dependence 
of the disk size, effective temperature, luminosity, and spectrum on the black hole mass, 
mass accretion rate, and disk viscosity.
                                                                                    
\subsection{Conductive Cooling-Dominant Corona}
                                
The condensation from a corona where cooling results mainly from conduction has 
been studied in earlier works by Liu et al. (2006) and Meyer et al. (2007). 
We introduce the main results here for completeness and for comparison to the 
case where the cooling by the inverse Compton scattering process is dominant 
(see \S 2.3). In the bulk of the corona the thermal state of the ions is not 
significantly affected by conductive cooling of the electrons until the electron 
temperature has become low enough near the base of the corona so that collisional 
coupling between the ions and electrons becomes effective. From this coupling 
interface down to the upper layers of the cool disk the ion temperature, 
$T_i$, and electron temperature, $T_e$, no longer differ. This results in a dramatic 
decline of $T_i$, leading to a significant increase in density as the pressure in 
the vertical extent changes little in comparison to the change in temperature. As a 
consequence, bremsstrahlung energy losses become much more important in this layer 
than in a typical ADAF. Henceforth, this layer is referred to as the radiating 
layer. The pressure in the ADAF is of particular interest since it determines 
whether evaporation or condensation takes place. For example, at sufficiently 
high pressure, bremsstrahlung can be so efficient 
that not only is all the heat drained from the ADAF by thermal conduction radiated 
away, but also the gas in the radiating layer is  further cooled, leading to condensation 
onto the disk.  On the other hand, if the pressure in the upper corona is 
sufficiently low so
that the density in the radiating layer is too low to radiate away the conductive 
flux, disk matter is  evaporated into the corona.  We adopt for our analysis the values derived by 
                       Narayan \& Yi (1995b). The ADAF pressure, density, 
viscous heating rate, and sound speed depend on the viscosity parameter, $\alpha$, black 
hole mass, $m$, the accretion rate, $\dot m$, and $r$ the 
distance from the black hole in the form (see Narayan \& Yi 
1995b),
\begin{eqnarray}\label{scaled}
p&=&1.87\times 10^{16}\alpha^{-1}m^{-1}\dot m r^{-5/2}
\, \rm{g cm^{-1} s^{-2}} \nonumber ,\\
n_e& =&5.91\times10^{19}\alpha^{-1}m^{-1}\dot m r^{-3/2} \,  \rm{cm^{-3}} \\
q^+&=& 2.24\times10^{20}m^{-2}\dot m r^{-4}\, \rm{ergs cm^{-3} s^{-1}}
\nonumber,\\
c_{\rm s}^2&=&1.67\times 10^{20} r^{-1} \,  \rm{cm^2 \,s^{-2}} \nonumber,
\end{eqnarray}
where $m$, $\dot m$ and $r$ are in units of solar mass, Eddington rate ($\dot M_{\rm Edd}
=1.39\times 10^{18}m$ g/s), and Schwarzschild radius, respectively.
                                                                                    
The ion number density is $n_i=n_e/1.077$, and the ion and electron temperatures closely 
follow
\begin{equation}
T_i+1.077 T_e =1.98\times10^{12} r^{-1}\rm K.
\end{equation}
The energy transfer from ions to electrons is given by Stepney (1983) and is approximated 
for the two-temperature advection-dominated hot flow  (Liu et al. 2002) as
\begin{eqnarray}\label {qie}
q_{ie} & = & 3.59\times 10^{-32} {\rm{g cm^{5} s^{-3} deg^{-1}}}
\, n_e n_i T_i
{\left(\frac{k T_e}{m_e c^2}\right)}^{-3/2}
  \nonumber \\
       & = & 1.05\times 10^{35}
T_e^{-3/2} \alpha^{-2}m^{-2} \dot m^{2}r^{-4} {\, \rm{ g cm^{-1}s^{-3}deg^{3/2}}}.  
\end{eqnarray}
The coupling temperature is reached when viscous 
            and compressive heating are balanced by the transfer of heat 
            from the ions to the electrons, $q_{ie}=q^+ + q^c$, which 
yields 
 \begin{equation}\label{T_cpl}
T_{\rm {cpl}}=1.98\times 10^9 \alpha^{-4/3}\dot m^{2/3}.
\end{equation}
The heat flux from the typical corona/ADAF to the radiating layer is derived from 
the energy balance in the corona, $\frac{dF_{\rm c}}{dz}=q_{ie}(T_e)$ assuming that 
the conductive flux is given by the expression $F_{\rm c}=-\kappa_0 T_e^{5/2}dT_e/dz$ (Spitzer 1962) with $\kappa_0 = 10^{-6}{\rm erg\,s^{-1}cm^{-1}K^{-7/2}}$.  (This value might be lower if e.g. chaotic magnetic fields 
 reduce the effective conductivity). This holds when the
collisional mean free paths are small compared to the 
length over which the electron temperature changes.
Solving for the conductive flux yields 
\begin{equation}\label{F_c^2}
F_{\rm c}^2(T_e)=(\kappa_0 K n_i n_e T_i)(T_{\rm em}^2-T_e^2)
\end{equation}
and at the coupling interface
\begin{equation}\label{Fc-general}
F_{\rm c}^{\rm ADAF}\approx -(\kappa_0 K n_i n_e T_i)^{1/2}T_{\rm em},
\end{equation}
where ``-'' means a downward directed heat flow, $K= 1.64\times 10^{-17} \rm {g cm^{5}s^{-3}deg^{1/2}}$, $T_{\rm em}$ is the 
maximum electron temperature at height $z_m$ corresponding to $F_c(z_m)=0$ and 
can be derived by integration of $F_{\rm c}=-\kappa_0 T_e^{5/2}dT_e/dz$ from $z_m$ 
to the interface by taking $F_c$ in Eq.\ref{F_c^2}, yielding 
\begin{equation}\label{T_em}
 T_{\rm em}=2.01\times 10^{10}\alpha^{-2/5}\dot m^{2/5} r^{-2/5}\rm K.
\end{equation}
The conductive flux from the corona arriving at the interface of the 
            radiation region is then
\begin{equation}\label{Fc-cond}
F_{\rm c}^{\rm ADAF}=-6.52\times 10^{24} \alpha^{-7/5}m^{-1}\dot m^{7/5} r^{-12/5}.
\end{equation}
The energy balance in the radiating layer is determined by the incoming conductive flux, 
bremsstrahlung radiation flux, and the enthalpy flux carried 
            by the mass evaporation/condensation flow,
\begin{equation}\label{energy}
\frac{d}{dz} \left[\dot m_z \frac{\gamma}{\gamma-1} \frac{1}{\beta}\frac{\Re T}{\mu} +
F_c \right] = -n_e n_i \Lambda(T).
\end{equation}
 This determines the  
evaporation/condensation rate per unit area, which  is  given by (Meyer et al. 2007)
\begin{equation}\label{cnd-general}
\dot m_z={\gamma-1\over \gamma} \beta{-F_{\rm c}^{\rm ADAF}\over {\Re T_{i}\over 
\mu_i}}\(1-\sqrt{C}\)
\end{equation}
with
\begin{equation}\label{C}
C \equiv\kappa{_0} b
\left(\frac{0.25\beta^2 p_0^2}{k^2}\right)
\left(\frac{T_{\rm {cpl}}}{F_c^{\rm{ADAF}}}\right)^2,
\end{equation}
where $b=10^{-26.56}\,\rm g\,cm^5s^{-3}deg^{-1/2}$ (Sutherland \& Dopita 1993), $\beta$ corresponds to the ratio of gas pressure to total pressure, $p_0={2\over \sqrt{\pi}}p$ the pressure in the radiating layer, $\gamma={8-3\beta\over 6-3\beta}$ (Esin 1997), and $\mu_i=1.23$ for 
an assumed chemical abundance of $X=0.75$, $Y=0.25$. 
Using  Eqs.\ref{T_cpl} and \ref{Fc-cond} for 
  $T_{\rm cpl}$ and $F_c^{\rm ADAF}$ and taking $\beta = 0.8$, 
            we obtain
\begin{equation}
C=0.96\alpha^{-28/15}\dot  m^{8/15}r^{-1/5}.
\end{equation}
For $C<1$, mass evaporates from the disk to the corona ($\dot m_z > 0$) since the 
conduction flux is not completely radiated away. On the other hand,  for $C>1$ 
coronal matter condenses into the disk ($\dot m_z < 0$) due to effective cooling 
by bremsstrahlung.  The condition, $C=1$, separates the regions of evaporation and 
condensation and, hence determines, the outer radius 
of the inner disk, given as  
\begin{eqnarray}\label{r_cnd1}
r_{\rm d}&=&0.815\alpha^{-28/3}\dot  m^{8/3}\nonumber\\
&=&5864\({\alpha\over 0.2}\)^{-28/3}\({\dot  m\over 0.1}\)^{8/3}.
\end{eqnarray}
Thus, the determination of the condensation/evaporation region depends on mass accretion rate and viscosity parameter in the ADAF. The integrated condensation rate (in units of 
the Eddington rate) from $r_{\rm d}$ to any radius ($r_i$) of the disk is
\begin{equation}\label{condensation_cnd}
\dot m_{\rm cnd}=\int_{R_i}^{R_d} {4\pi R \over \dot M_{\rm Edd}}\dot m_zdR= 3.23\times 10^{-3}\alpha^{-7}\dot m^3 f(r_{\rm i}/r_d),
\footnote{Eq.(29) in Meyer et al. 
    (2007) contains an error, without consequences in the paper since 
    the formula was not further used. The true value is $3.23\times 10^{-3}$ as here}
\end{equation}
with
\begin{equation}
f(x)=1-6x^{1/2}+5x^{3/5}.
\end{equation}

\begin{figure}
\plotone{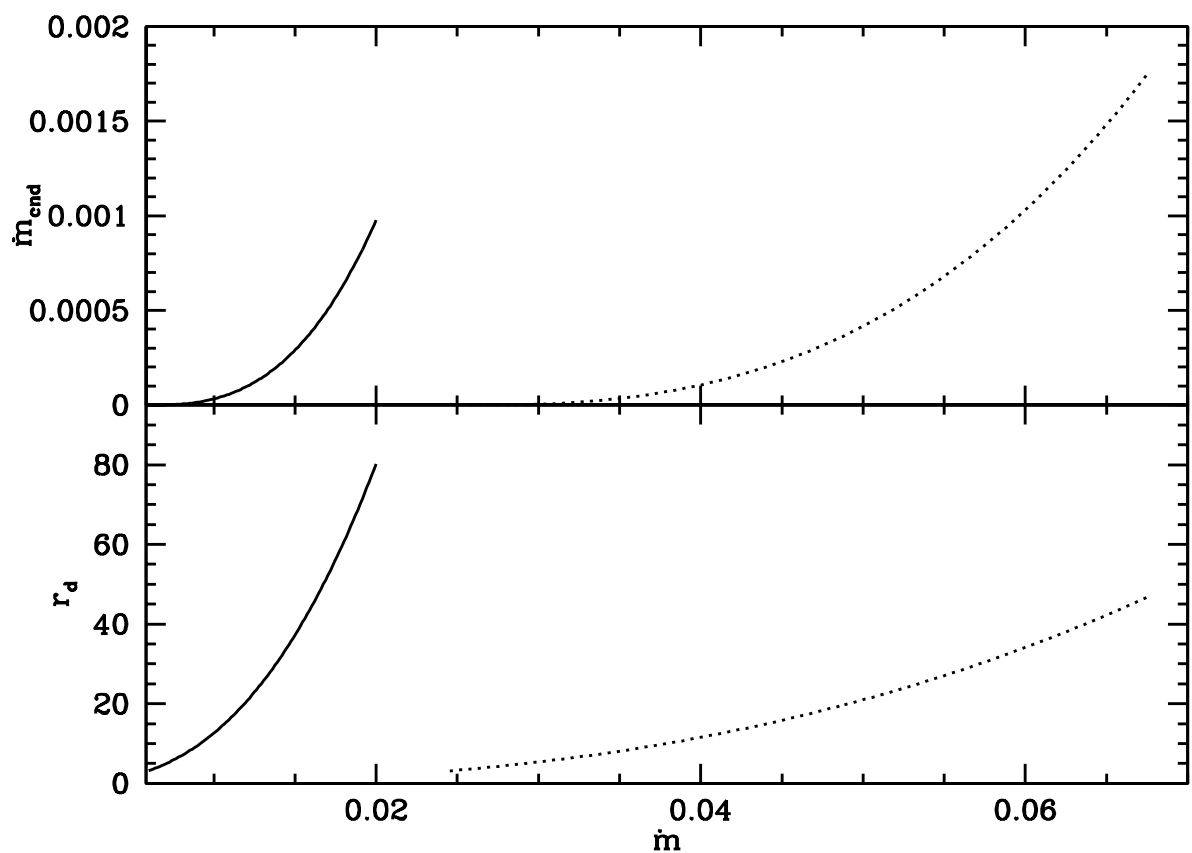}
\caption{The dependence of the critical condensation radius ($r_d$) and the 
condensation rate ($\dot m_{\rm cnd}$, integrated from $r_d$ to  $r_{\rm 
tmax}$) on the accretion rate ($\dot m$) for a conductive cooling-dominant 
corona. The mass of black hole is fixed to $10M_\odot$. The values of the 
viscosity  are $\alpha=0.2$ (solid curves) and $\alpha=0.3$ (dotted curves), 
respectively.}
\label{f:r_d-cnd}
\end{figure}

The expressions for the size of the inner disk, $r_d$ (eq.\ref{r_cnd1})
and the mass condensation rate, $\dot m_{\rm cnd}$ (eq. \ref{condensation_cnd})
reveal their sensitivity to the accretion rate and viscosity. Their dependence  
on the accretion rate ($\dot m$) for $\alpha=0.2$ and $\alpha=0.3$ are shown 
in Fig.\ref{f:r_d-cnd}. It can be seen that an inner disk with size $r_d<100$ 
only exists in a limited range of accretion rates, and this range strongly depends 
on $\alpha$.

The radiation from the corona is dominated by bremsstrahlung in the radiating layer, 
$$F_{\rm Brem}=\int_{z0}^{z1}n_e n_i \Lambda(T)dz,$$
 where $z0$  and $z1$ are the lower and upper boundary of the radiating layer. By 
combining Eqs.\ref{energy}, \ref{cnd-general} and \ref{C},  the flux is given by 
\begin{equation}
F_{\rm Brem}=-F_c^{\rm ADAF}\sqrt{C}=6.391\times 10^{24} \alpha^{-7/3}m^{-1}\dot m^{5/3}r^{-5/2}\,
\rm erg\ s^{-1}\,cm^{-2}
\end{equation}
and the corresponding luminosity from the two sides of the disk corona  is
\begin{eqnarray}\label{L_cnd}
{L_{\rm Brem}\over L_{\rm Edd}}&=&2\int_{3R_S}^{R_d} 2\pi R F_{\rm Brem}dR=
0.0642 \alpha^{-7/3}\dot m^{5/3}\[1-\({3\over r_d}\)^{1/2}\]\nonumber\\
&=&0.0591\({\alpha\over 0.2}\)^{-7/3}\({\dot m\over 0.1}\)^{5/3}\[1-\({3\over r_d}\)^{1/2}\].
\end{eqnarray}

\subsection{Compton Cooling-Dominant Corona}

In the regime at relatively high mass accretion rates or with a luminous 
soft photon flux, the inverse Compton scattering process could be more 
effective for cooling the electrons than the vertical thermal conduction in an ADAF-like corona.  
The electron temperature in the upper coronal region then would be determined 
by Compton cooling. However, in the presence of an underlying cool disk 
conduction sets in at some height as the dominant  cooling mechanism for the 
lower coronal layers. The additional Compton cooling within the corona leads to a 
relatively cool electron component in the corona/ADAF, resulting in a conductive 
flux from the corona to the radiating layer lower than that estimated in \S 2.2.  The electron 
temperature in the corona is now determined by $q_{\rm ie}=q_{\rm cmp}$
where $q_{\rm cmp}$ is the Compton cooling rate given by 
\begin{equation}\label{q_comp}
q_{\rm cmp} = {4kT_e\over m_ec^2} n_e \sigma_T c u
\end{equation}
where $u$ is the soft photon energy density.  By assuming that the soft photons for Compton scattering arise 
from the local underlying disk, $u$ is expressed in terms of effective temperature in the local disk as 
$u(r)={1\over 2}aT_{\rm eff}^4(r)$. The factor ${1\over 2}$ for the 
energy density of an isotropic blackbody photon field is taken since photons from the 
underlying disk cover only half of the sky of electrons in the corona. The effective 
temperature of a steady state disk with a constant mass accretion rate, given by 
$\sigma T_{\rm eff}^4 ={3GM\dot M_{\rm d}\over 8\pi R^3}\[1-\({3R_S\over R}\)^{1/2}\]$, 
reaches at a maximum $T_{\rm eff, max}$ at $R=(49/36)\times 3R_S$ and is expressed as 
\begin{equation}
T_{\rm eff}(r)=2.05T_{\rm eff,max}\({3\over r}\)^{3/4}\[1-\({3\over r}\)^{1/2}\]^{1/4}
\end{equation} 
With this effective temperature for the soft photon field, the electron temperature of the corona is derived from Eq.\ref{q_comp} as 
\begin{equation}\label{T_ec}
 T_{\rm ec}=3.025\times 10^9\alpha^{-2/5} m^{-2/5}\dot m^{2/5}r^{1/5}\[1-\({3\over r}\)^{1/2}\]^{-2/5}\({T_{\rm eff,max}\over 0.3keV}\)^{-8/5}\rm K,
\end{equation}
  This temperature, $T_{\rm ec}$, is 
characteristic of the upper Compton cooling dominant corona and represents 
the maximum temperature within a given column at a distance, $r$. 
The conductive flux from the corona to the coupling interface is calculated 
from Eq.\ref{Fc-general} by replacing $T_{\rm em}$ with $T_{\rm ec}$ and is 
given by
\begin{equation}\label{Fc-comp}
F_c^{\rm ADAF}=-9.816\times 10^{23}\alpha^{-7/5} m^{-7/5}\dot m^{7/5}r^{-9/5}\[1-\({3\over r}\)^{1/2}\]^{-2/5}
\({T_{\rm eff,max}\over 0.3keV}\)^{-8/5}
\rm erg\,s^{-1}\,cm^{-2},
\end{equation}
Since the Compton cooling rate is much lower than the bremsstrahlung rate 
in the radiating layer, the energy balance and, hence, the 
evaporation/condensation rate is the same as in the conduction dominant 
cooling case, except for the fact that the conductive flux expressed in 
Eq.\ref{cnd-general} is replaced by Eq.\ref{Fc-comp}. With this expression 
for the conductive flux, the critical condensation radius and the integrated 
condensation rate are derived by setting $C=1$ (in Eq.\ref{C}) and by 
integrating Eq.\ref{cnd-general} respectively, yielding
\begin{equation}\label{r_cnd2}
r_{\rm d}\[1-\({3\over r_d}\)^{1/2}\]^{-4/7}=14.417\alpha^{-4/3}m^{4/7}\dot m^{8/21}\({T_{\rm eff,max}\over 0.3keV}\)^{16/7},
\end{equation}
and
\begin{equation}\label{condensation-cmp}
\dot m_{\rm cnd}(r)=A\left\{2B\[\({r_{\rm d}\over r}\)^{1/2}-1\]-\int_{r/3}^{r_{\rm d}/3}x^{1/5}\(1-x^{-1/2}\)^{-2/5}dx \right\}
\end{equation}
where
\begin{equation}\label{A}
A=6.164 \times 10^{-3}\alpha^{-7/5}m^{-2/5}\dot m^{7/5}\({T_{\rm eff, max}\over 0.3keV}\)^{-8/5}
\end{equation}
and
\begin{equation}\label{B}
B=3.001\alpha^{ -14/15}m^{2/5}\dot m^{ 4/15}\({T_{\rm eff, max}\over 0.3keV}\)^{8/5}\({r\over 3}\)^{1/2}.
\end{equation}
 There is either no solution or two solutions for $r_d$ in Eq.\ref{r_cnd2}, depending on system parameters. In the case of no solution,  mass does not condense into the disk by Compton cooling, 
but rather evaporates from the disk to the corona. The two solutions $r_{d1}$ and $r_{d2}$ 
determined by Eq.\ref{r_cnd2} lie at both sides of the distance $r=(81/49)\times 3$, 
yielding a spatial extent for the condensation region described by $r_{d1}<r<r_{d2}$.

The total condensation rate to the inner disk is the integral from $r_{\rm d2}$ to 
$r_{\rm d1}$. Since the accretion rate in the disk increases with decreasing distance 
until $r=r_{d1}$, the disk effective temperature does not reach its maximum at 
$r_{\rm tmax}=(49/36)\times 3$, but at some distance slightly smaller than it (depending 
on the system parameters). However, the true maximal effective temperature is not 
much larger than the value at $r_{\rm tmax}$.
In the following we assume the maximal effective temperature is reached at $r_{\rm tmax}$, 
which is a good approximation. 

\begin{figure}
\plotone{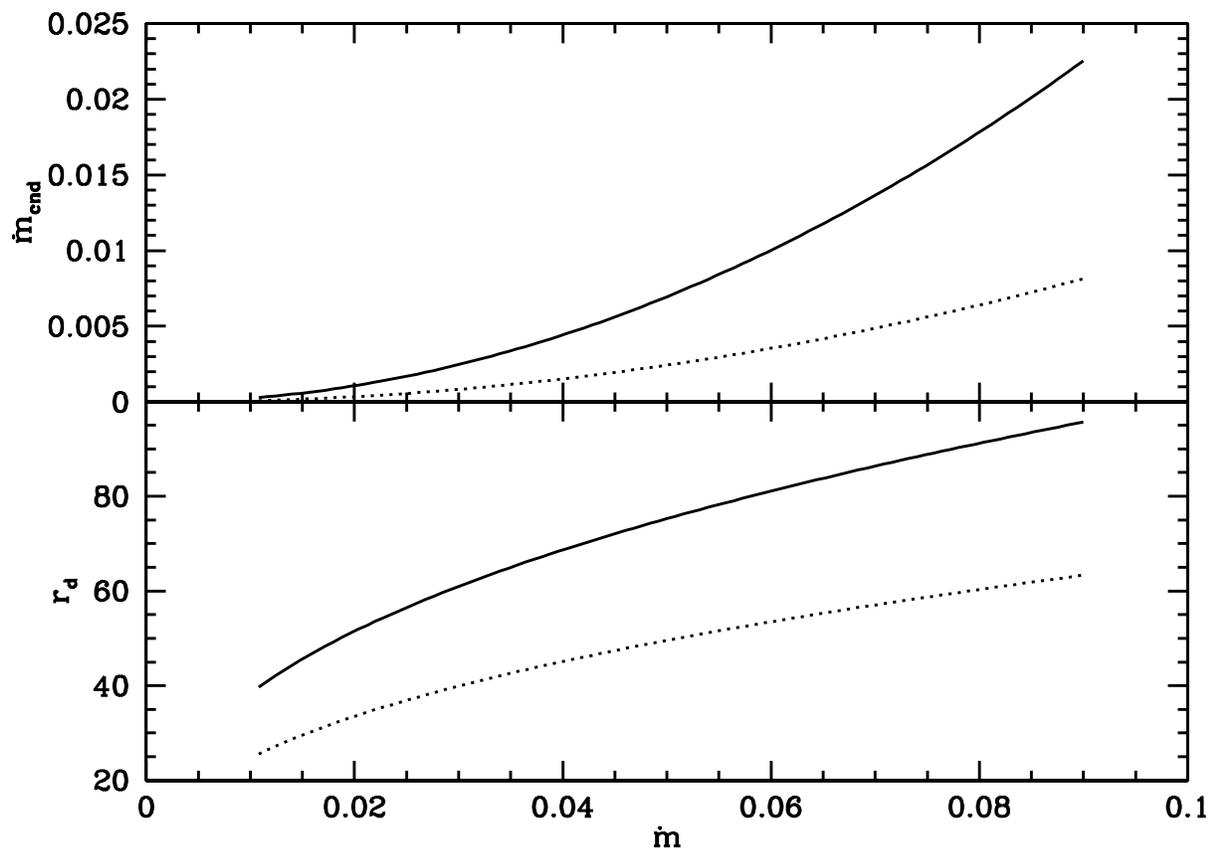}
\caption{The dependence of the critical condensation radius ($r_d$) and the 
condensation rate ($\dot m_{\rm cnd}$, integrated from $r_d$ to $r_{\rm 
tmax}$) on the accretion rate ($\dot m$) for a Compton cooling-dominant corona. 
The mass of black hole is fixed to $10M_\odot$ and the effective temperature 
of the disk $T_{\rm eff, max}=0.3$keV. The values of the viscosity are $\alpha=0.3$ 
(solid curves) and $\alpha=0.4$ (dotted curves), respectively.}
\label{f:r_d-cmp}
\end{figure}

Thus, for a Compton cooling dominant corona, the region where matter condenses
from the corona to the disk and the total condensation rate are determined by 
the mass of the black hole, the mass accretion rate, effective temperature of 
the soft photon radiation, and the disk viscosity parameter.   Fig.\ref{f:r_d-cmp} 
shows that for given black hole mass, $m=10$, and disk temperature, $T_{\rm eff, 
max}=0.3keV$, the size of the inner disk and the condensation rate increase 
with accretion rate.  For larger $\alpha$ values, both the size of the inner disk 
and the mass condensation rate decrease for a given $m$ and $T_{\rm eff, max}$.

If the effective temperature is assumed to originate from disk accretion fed 
by condensation, the maximal effective temperature at $r_{\rm tmax}$ 
is no longer a free parameter, 
\begin{eqnarray}\label{T_eff}
T_{\rm eff, max}&=&1.3348\times 10^7 \,{\rm K} m^{-1/4}\dot m_{\rm cnd}^{1/4}(r_{\rm tmax}) \nonumber \\
&=&0.2046{\rm keV}\({m\over 10}\)^{-1/4}\[{\dot m_{\rm cnd}(r_{\rm tmax})\over 0.01}\]^{1/4}.
\end{eqnarray}
Replacing $T_{\rm eff, max}$ in Eqs.\ref{r_cnd2}, \ref{A} and \ref{B}, 
Eq.\ref{condensation-cmp} becomes a non-linear equation for $\dot 
m_{\rm cnd}(r_{\rm tmax})$ and can be numerically solved for given $\alpha$, $m$ and 
$\dot m$. The integrated condensation rate at any distance $r$ can then also be 
calculated for a known $T_{\rm eff,max}$ using Eq.\ref{T_eff}.  

Finally, the luminosity associated with the inverse Compton scattering process 
in the corona can be calculated as 
\begin{equation}
 L_{\rm cmp}=\int_{R_{\rm d1}}^{R_{\rm d2}} 2\pi R H {4kT_e\over m_e c^2} n_e \sigma_T 4\sigma T_{\rm eff}^4 (R) dR.
\end{equation}
Replacing the temperature $T_e$ by Eq.\ref{T_ec} and the density by 
Eq.\ref{scaled}, we have
\begin{equation}\label{L_cmp}
{L_{\rm cmp}\over L_{\rm Edd}}=0.392\alpha^{-7/5}m^{3/5}\dot m^{7/5}\({T_{\rm eff, max}\over 0.3keV}\)^{12/5}\int_{r_{\rm d1}/3}^{r_{\rm d2}/3}x^{-23/10}\(1-x^{-1/2}\)^{3/5}dx.
\end{equation}

\subsection{Compton Dominant or Conduction Dominant-Cooling?}

\begin{figure}
\plotone{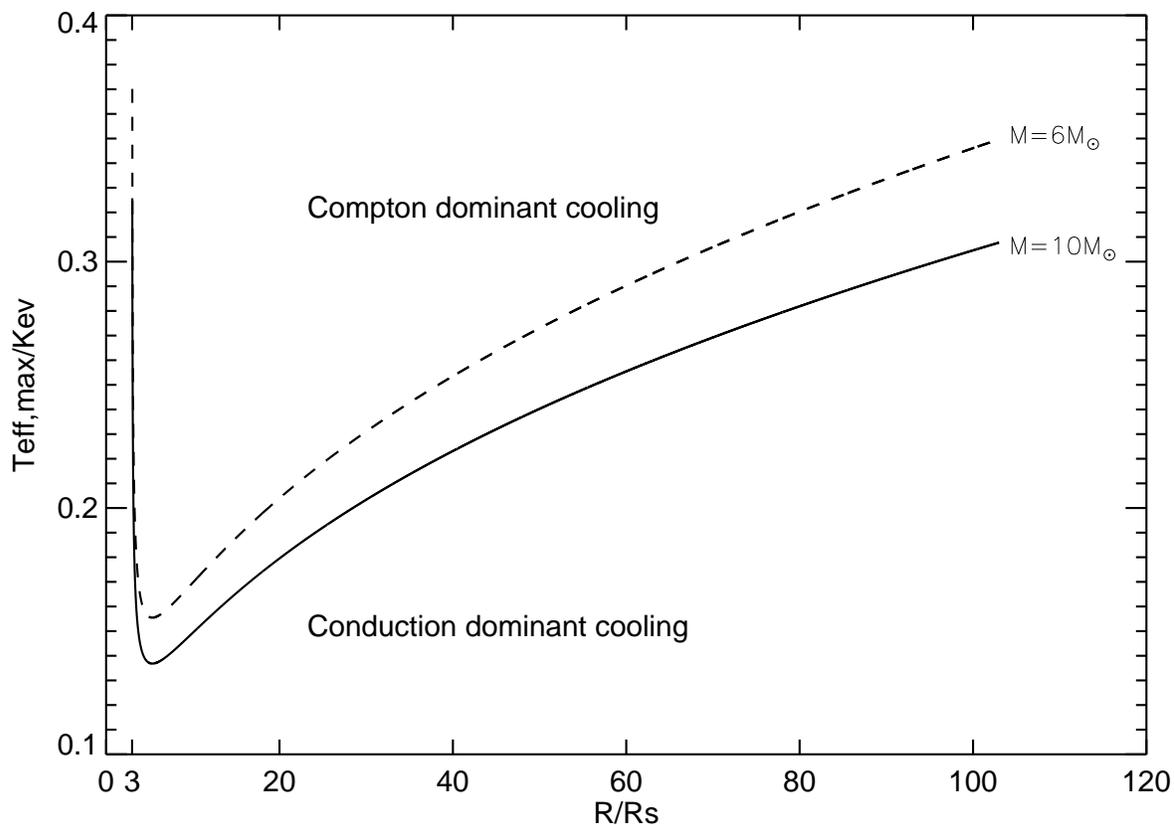}
\caption{The radial region where Compton or conduction cooling dominates as a 
function of the innermost disk temperature. The curves represent the critical 
effective temperature of the inner disk, above which Compton cooling should 
be taken into account and below which the conduction dominant model is valid.  
The solid and dash lines correspond to a black hole mass of $m=10$ and $m=6$ 
respectively.}
\label{f:cmp-cnd}
\end{figure}

In general, cooling from both the Compton scattering and thermal conduction 
processes should be included, however, in order to obtain analytical results 
we have only considered  the cases when one of these processes is dominant. 
In the 
following we delineate the physical regime in which Compton cooling is 
important.                                                                                     
One approach for determining the importance of Compton and conduction cooling 
in the corona is to compare the electron temperatures corresponding
to the Compton cooling and conductive cooling regimes. The process resulting 
in a lower temperature is the more efficient cooling mechanism and thus would 
be dominant.  As shown in previous work (e.g. Meyer et al. 2007), thermal 
conduction results in a vertical temperature distribution in the corona with 
the electron temperature decreasing from a maximum value $T_{\rm em}$ (see 
Eq.\ref{T_em}) at the highest layer $z_m$ to the coupling temperature $T_{\rm cpl}$ 
at the interface between the corona and radiating layer. On the other hand, for 
inverse Compton scattering, the electrons in the ADAF-like corona cool to a 
temperature  $T_{\rm ec}$, which is independent of the vertical height. If, at 
any given height $z<z_m$, the electron temperature $T_e(z)$ determined by 
conduction is larger than the Compton cooling temperature $T_{\rm ec}$, 
Compton cooling should be more important.  In this case, the maximum temperature 
$T_{\rm em}$ of  Eq.\ref{T_em} is no longer relevant and the temperature from $z$ 
to $z_m$ is determined by Compton cooling.  That is, $T_e=T_{\rm ec}$, which 
becomes the maximum temperature in the upper coronal region. As long as the 
Compton temperature $T_{\rm ec}$ is less than the maximum temperature $T_{\rm 
em}$ as determined by conduction, Compton scattering should contribute to the 
coronal cooling.  From $T_{\rm ec}\le T_{\rm em}$, the critical radius where 
Compton cooling sets in is given by
\begin{equation}\label{r_cmp}
r_{\rm cmp}\[{1-\({3\over r_{\rm cmp}}\)^{1/2}}\]^{-2/3}\le 23.487m^{2/3}\({T_{\rm eff,max}\over 0.3keV}\)^{8/3}.
\end{equation}
There exists a minimum  at $r=(16/9)\times 3$ in the  function for $r_{\rm cmp}$ on left-hand side of  Eq.\ref{r_cmp}.  Thus, for this equation there is either no solution, implying 
that Compton cooling is unimportant or two solutions exist interior and 
exterior to $r=(16/9)\times 3$, which determines the Compton dominant region, 
$r_{\rm cmp1}<r<r_{\rm cmp2}$. 
This indicates that Compton cooling is non-negligible in the inner region 
($r\sim 16/3$) for X-ray binaries if the disk effective temperature is not very low.
For given $M$ and $T_{\rm eff, max}$, say,   $M=10M_\odot$  and $T_{\rm eff,max}=0.3$keV,  the Compton dominant region  $3.03<r<96$ follows, whereas for  
$M=10M_\odot$ and $T_{\rm eff, max}= 0.2$keV, a smaller region $3.16<r<28.5$ follows.  
Fig.\ref{f:cmp-cnd} illustrates the radial extent where Compton cooling or conductive 
cooling is important in terms of the disk effective temperature in the innermost 
region for black hole masses of $m=10$ and $m=6$.  It can be seen that, for 
typical X-ray binaries, the Compton cooling plays a role  if the effective 
temperature in the inner disk is higher than about 0.14 keV.

\subsection{Caveats}

We wish to point out that the dependence of the ADAF solutions on the viscosity 
parameter $\alpha$ not only appears explicitly via its power-law dependence, as 
shown in Eq.\ref{scaled}, but also implicitly through the coefficients of these 
solutions (Narayan \& Yi 1995b). The values of the coefficients  in Eq.
\ref{scaled} are derived by assuming $\alpha=0.2$. As this dependence is weak, 
we neglect the implicit dependence in the following analysis when $\alpha$ is 
varied.  Similarly, the coefficients in the ADAF solution also implicitly depend 
on the parameterized magnetic field $\beta$. In this work $\beta=0.8$ is assumed 
based on results of the MHD simulation of Sharma et al. (2006). Compressive 
heating in the ADAF ($q^c={1\over 1-\beta}q^+$) is also calculated for this value. 
As remarked earlier, the influence of the magnetic field on the vertical conduction 
is not discussed in our present analysis.
Furthermore, the effects of irradiation and disk reflection are assumed to be 
unimportant since it is not expected to be significant for the small size of 
the weakly emitting inner disk that is characteristic for the low/hard state 
although it can lead to additional heating of the disk (see \S 5).  Finally, 
we do not introduce
a color correction factor to describe the spectral hardening factor of the disk 
in our simple model since such a description is not an adequate description of 
the spectrum at the low mass accretion rate levels (see Shimura \& Takahara 1995) 
in the inner disk of our systems in their low hard state. 

\section{Results}

For a system with mass accretion rates lower than the maximum evaporation rate, 
corresponding to the hard-soft state transition, a gap forms in the disk by the 
efficient evaporation of matter at about a few hundred Schwarzschild radii. An
inner cool remnant disk can be present extending from the ISCO to the critical 
condensation radius $r_{\rm d}$. This inner disk is fed by the condensing 
matter from its overlying corona and the accretion rate in the cool disk represents 
the integrated condensation rate.  The coexistence of this inner cool disk and 
a coronal gap can yield an intermediate state spectrum characterized by 
both hard and soft components. Here, the strength of the soft component depends 
on the condensation rate and the inner disk size, both of 
which are determined by the system parameters, i.e. the mass of the black hole, 
the mass accretion rate, and the viscosity parameter.

\subsection{Illustrative Examples for Conduction Dominant Cooling}

For a typical X-ray binary with $M=10M_\odot$ and standard disk viscosity parameter, 
$\alpha=0.2$, the accretion rate should be less than the corresponding transition 
rate, 0.02, if the system is in the low or intermediate state. As an example, we adopt 
a value for the mass accretion rate of $\dot m=0.01$.  Based on the analysis 
presented in Sect.2.2 for the conduction dominant cooling model, a critical 
condensation radius of $r_d=12.6$, a total condensation rate of $\dot m_{\rm 
cnd}=4.70\times 10^{-5}$, an inner disk effective temperature of $T_{\rm eff, max}$
=0.05keV, and a luminosity of $L/L_{\rm Edd}=6.53\times 10^{-4}$ are found. To 
check the consistency of the solution, the critical radius for the Compton 
cooling region was calculated.  A Compton dominant cooling region was 
not found to exist, implying that the conduction dominant model is consistent.  
For lower mass accretion rates, $\dot m< 0.01$ or larger coronal viscosity 
parameters, $\alpha > 0.2$, both the disk size and the 
mass condensation rate rapidly decrease. Although the conductive cooling model 
remains a valid description, the inner disk vanishes for $\dot m<0.006$ (for 
$\alpha=0.2$) or $\alpha>0.233$ (for $\dot m=0.01$).  For higher mass accretion 
rates ($\dot m>0.01$) or smaller viscosity parameters, the condensation 
radius and mass condensation rate increase, leading to higher disk temperatures. 
Only for sufficiently high mass accretion rates (e.g. $\dot m=0.03$ but standard 
viscosity, $\alpha=0.2$), or sufficiently small viscosities (e.g. $\alpha=0.14$, but 
unchanged accretion rate, $\dot m=0.01$), must Compton cooling be taken into account
in describing the physical state of the inner region.

\subsection{Illustrative Examples for Compton Dominant Cooling}\label{s:cmp}

To illustrate an X-ray binary where Compton cooling plays an 
important role, we consider a system with a black hole mass of $M=10M_\odot$ and 
a  viscosity parameter $\alpha=0.35$, but assume a higher mass accretion 
rate $\dot m=0.11$.  As discussed in \S 3.1, Compton cooling is expected 
to be an important cooling process in the inner region. Calculations based 
on conduction alone would yield condensation from the ISCO up to $r=41$, resulting
in a condensation rate of $2.79\times 10^{-3}$ times 
the Eddington rate. This leads to a maximal effective temperature of the 
inner disk of 0.14 keV (Eq.\ref{T_eff}). At such a temperature, Compton 
cooling becomes important from $r=4$ until $r=8$ (as calculated from 
Eq.\ref{r_cmp}).

Therefore, in the case of  $\alpha=0.35$ and  $\dot m=0.11$, cooling in the corona is 
dominated by the Compton scattering in the inner region and the vertical conduction 
in the surrounding region.  By presuming a value for the disk effective temperature, 
we calculate the condensation rate from Eqs.\ref{r_cnd2}-\ref{B} and derive the 
effective temperature from Eq.\ref{T_eff}. If this temperature matches the 
presumed value, the solution for the condensation rate is self-consistent.  
By iteration we find a consistent solution for the total condensation 
rate, $\dot m_{\rm cnd}=3.0\times 10^{-3}$, obtained by integration from an inner Compton 
dominant region ($3.7<r<10.6$) and an outer conduction dominant region 
($10.6<r<41$), with a disk effective temperature $T_{\rm eff, max}$=0.15keV 
at $r_{\rm tmax}$.

In the above example, Compton cooling is not very strong, leading to a slight 
increase in the rate of condensation.  For very efficient Compton cooling, the 
decrease in the conductive flux results in a significant increase in the 
condensation rate. To illustrate such a case we assume $\alpha=0.4$, $\dot m=0.16$, 
and $m=10$.  For this set of parameters, calculations based on the Compton cooling 
model show that the coronal gas condenses into the disk from $r=30.5$ to $r=3.08$, 
resulting in a mass condensation rate of $\dot m_{\rm cnd}=0.01$. Such a rate leads 
to a consistent effective temperature of $T_{\rm eff, max}=0.205$keV at $r_{\rm tmax}$. 
On the other hand, calculation based on the conduction model shows a condensation rate of only 0.003, much smaller than predicted by Compton cooling. Further calculations 
of the Compton dominant region show that the Compton scattering dominates the coronal 
cooling nearly throughout the condensation-fed inner disk.

Examples for various values of the viscosity and accretion rate are listed 
in Table~\ref{t:examples}. It is shown that the conduction cooling leads to low 
condensation rates and hence low disk temperatures, while the Compton cooling 
results in high condensation rates and hence high effective temperatures. For 
comparable viscosity values, the Compton cooling becomes efficient at high 
accretion rates.

\begin{figure}
\plotone{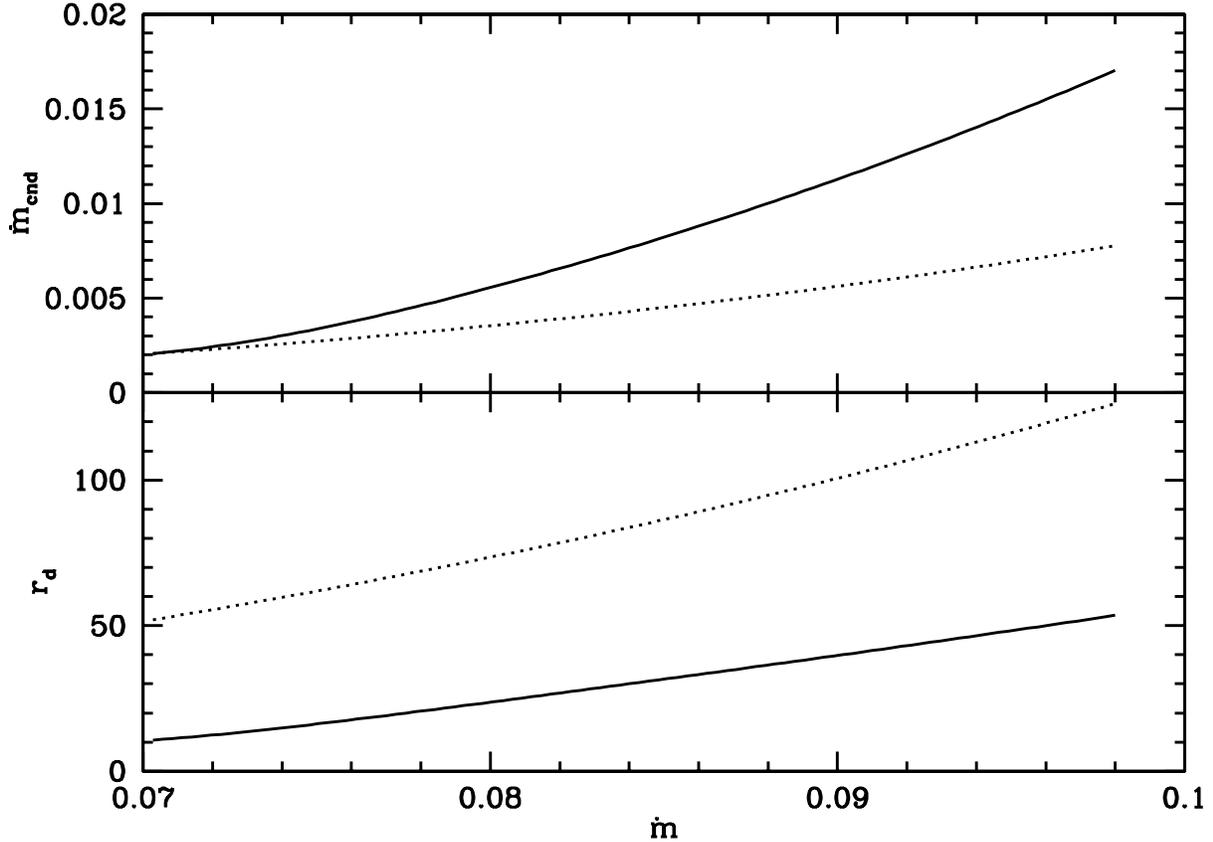}
\caption{Comparison of the critical condensation radius ($r_d$) and the integrated 
condensation rate ($\dot m_{\rm cnd}$) for a conductive cooling and Compton 
cooling-dominant corona. The mass of black hole is fixed to $10M_\odot$, and the 
value of the viscosity is 0.3. The dotted curves refer to a conductive cooling 
corona as shown in Fig.\ref{f:r_d-cnd}. The solid curves are for a Compton cooling 
dominant corona, where $\dot m_{\rm cnd}$ is integrated from an inner Compton 
dominant to an outer conductive dominant region. The smaller critical radius 
resulting from Compton cooling reflects the fact that Compton scattering is 
important mainly in the inner region.  The higher condensation rate in 
the case of including Compton scattering also indicates that the Compton 
cooling can play an important role in condensation.}
\label{f:comparison}
\end{figure}

Neglecting the upper limit to the accretion rates, we show in Fig.\ref{f:comparison}
the dependence of the condensation rate on the mass accretion rate for the 
case of Compton dominant cooling in the inner corona and conduction dominant cooling 
in the outer corona. Condensation caused by conduction alone is also shown for 
comparison.  It can be seen that the condensation is strong for Compton 
cooling, though it only dominates over conduction in the inner region. 

\section{Comparison with Observations}

To make a meaningful comparison between observations and theory it is important
to recognize that the observations provide information on the luminosity rather 
than the mass accretion rate. In the case of an optically thick standard thin 
disk, the mass accretion rate is deduced from the luminosity, $\dot M= L/ \eta c^2$ 
by assuming an energy conversion efficiency of $\eta=0.1$.  However, in an 
ADAF-like corona, $\eta$ is less than 0.1 and its variation depends on the 
structural disk parameters ($p$, $T_e$, etc), and therefore is implicitly determined 
by $\alpha$, $m$, and $\dot m$.  For a Compton cooling dominant corona the main 
contribution to the luminosity is due to inverse Compton scattering. Bremsstrahlung 
radiation from the radiation layer could also contribute, but as a small part. Therefore, 
with observationally inferred mass and effective temperature, a value for viscosity, and 
a tentative $\dot m$, we calculate numerically the radiative luminosity from both Compton 
scattering (Eq.\ref{L_cmp}) and Bremsstrahlung (Eq.\ref{L_cnd}) . If the derived 
luminosity is not equal to the observed value, the accretion rate is varied 
and the calculation repeated until a consistent luminosity is obtained. In this way, the 
accretion rate can be found.  The crucial issue in the modeling is whether a value 
for $\dot m$ exists which predicts condensation rather than evaporation. 

In the case of a corona dominated by thermal conduction, the luminosity is 
due to bremsstrahlung radiation, and the mass accretion rate only depends 
on the observed luminosity and the assumed $\alpha$ in the form
\begin{equation}\label{mdot-B}
\dot m=5.196\alpha^{7/5}\({L_{\rm Brem}\over L_{\rm Edd}}\)^{3/5}
\[1-\({3\over r_d}\)^{1/2}\]^{-3/5},
\end{equation}
where the the factor $\[1-\({3\over r_d}\)^{1/2}\]^{-3/5}$ is approximately unity
as long as $ r_d$ is not very small.  The inner disk size $r_d$, the condensation 
rate, and the effective temperature are thus determined from Eq.\ref{r_cnd1}, 
Eq.\ref{condensation_cnd} and Eq.\ref{T_eff} respectively. 
We note that the conduction model is based on the assumption that Compton cooling 
is negligible and would only be consistent with the observations for condensation rates
yielding low disk temperatures (from Eq.\ref{T_eff}).

A constraint on the viscosity parameter can  be obtained from the 
mass accretion rate corresponding to the transition between spectral states. 
As shown in previous work by Meyer et al. (2000a), the maximum evaporation 
rate, representing the hard-soft transition rate, roughly depends on the value of 
the viscosity parameter as $\dot m_{\rm tr}\propto \alpha^3$.  In the intermediate 
state discussed here, the mass accretion rate cannot be larger than this 
transition rate for, otherwise, the optically thick disk extends to the ISCO 
without a coronal gap, resulting in a soft thermal dominated spectrum. 

\subsection{GX339-4}

Based on the optical spectroscopy of GX339-4 during outburst, Hynes et al. (2003) 
infer an orbital period of 1.7557 days and a mass function of $5.8 \msun$ for the 
system.  For a likely distance  we take 8 kpc and for the black hole mass $10 \msun$
(see Zdziarski et al. 2004).  Recently, Miller et al. (2006b) observed GX 339-4 
during the low hard state with XMM Newton and found a soft component of $kT=0.3$keV, 
interpreting it to arise from an inner disk extending to the ISCO. Such a weak disk 
near the black hole is very difficult to explain within the framework of standard
accretion disk models. As the density of the reflecting medium inferred 
from observations is low (Miller et al. 2006b), the inner disk must be 
spatially limited in extent. At these low temperatures, in contrast to the 1-2keV 
spectrum in the soft state, a small inner disk would also be necessary in order that 
strong irradiation by the corona can be avoided. Given an unabsorbed flux of $5.33\times 
10^{-9} {\rm ergs\,s^{-1}cm^{-2}}$ (Miller et al. 2006b), the luminosity is $L/L_{\rm Edd}=
0.03$.  At this luminosity and the characteristic power-law spectrum of GX 339-4 in the 
low hard state, Compton scattering could be dominant.  Upon comparing the electron 
temperature obtained from analysis for the two cooling mechanisms, we confirm that Compton 
cooling is important  from the innermost region  $r=3.03$ up to $r_{\rm cmp}=95.7$ (by Eq.\ref{r_cmp}).  Even if a colour correction of 1.7 (Shimura \& Takahara 1995) is assumed, the Compton cooling is still dominant in the inner region to the distance of $r=18$. Hence, we consider GX 339-4 in terms of the Compton dominant condensation model.

With the black hole mass, luminosity, and effective temperature taken from the observations,  
and a standard value 0.2 for $\alpha$, it is found that the accretion rate is  $\dot m=0.024$.
This yields an inner disk extending to a distance of 99.5 Schwarzschild and a luminosity 
ratio of  21.8\% between the disk and the corona.  Adopting a larger viscous value, $\alpha=0.3$, 
one obtains an accretion rate of 0.037. The disk size is  $r_{\rm d}=66.5$ and disk 
luminosity fraction is 12.7\%, smaller than the case of standard viscosity value. A 
further increase in the viscosity to $\alpha=0.4$ leads to a smaller inner disk and a lower 
disk luminosity fraction. Detailed fitting results are presented in Table~\ref{t:fits}, where for 
a given value of $\alpha$, a value for $\dot m$ is found, which leads to a luminosity
consistent with the inferred value. The corresponding spatial extent of the condensation 
region, the condensation rate, the effective temperature, as well as the ratio of the disk 
to corona luminosities are also listed.

By comparing modeling results, we find that a larger viscosity predicts a smaller disk 
size and disk contribution to the luminosity. 
We note that the condensation rate yields an effective temperature (as shown in column 7 
in Table~\ref{t:fits}), which is lower than observed. This may indicate a need for an additional 
energy source above that associated with the accretion of condensed gas. For example, 
other processes, such as irradiation, may lead to additional heating of the gas in the 
disk surface. Notwithstanding the uncertainties in observationally determining 
the color temperature of the soft thermal component, the lack of agreement may also 
imply a need for a color correction factor, although, in that case, Compton cooling may 
not be as effective.
 
\subsection{J1753.5-0127}

We also can compare our model with the recent observations of J1753.5-0127 (Miller 
et al. 2006a), however there are significant uncertainties in the distance and 
mass of the black hole. Our comparisons, thus, should be only considered to be 
indicative and accurate modeling must await determinations of these parameters. 
XMM Newton observations of J\,1753.5-0127 reveal a X-ray luminosity of $L_X/L_{\rm Edd}= 
0.003(D/8.5kpc)^2(M/10M_\odot)$ in the range 0.5keV to 10keV and a hard X-ray 
spectral index of $\sim 0.66$ (Miller et al. 2006a).  Extrapolating the luminosity 
to 100-200keV yields ${L\over L_{\rm Edd}}\approx 0.01(D/8.5kpc)^2(M/10M_\odot)$. 
The fit to the X-ray spectrum reveals a cool disk component of $kT=0.2$keV 
(Miller et al. 2006a).  Taking $D=8.5 kpc$ and $M=10M_\odot$, the Compton 
process would be dominant in the inner region from $r=3.16$  to $r=28.5$ as estimated 
from Eq.\ref{r_cmp}.  Therefore, to model this object, Compton cooling should be 
taken into account. For the standard viscosity, $\alpha=0.2$, we find a mass 
accretion rate of $\dot m=0.02$ would suffice to produce the luminosity from 
both Bremsstrahlung in the radiating layer and Compton scattering in the typical coronal
region. The size of 
the  inner disk is $r_{\rm d}=33.4$  (from Eq.\ref{r_cnd2}), while from  Eq.\ref{r_cnd1} 
$r_{\rm d}=80$.  This implies that condensation begins at $r=80$ due to conduction, 
while Compton cooling sets in at $r_{\rm d}=28.5$ and dominates to $r_{\rm d}=3.16$.
In this case, the condensation rate is obtained from an integration from the outer 
conduction dominant region ($28.5\le r\le 80$) to the inner Compton dominant region 
($3.16\le r\le 28.5$), yielding $\dot m_{\rm cnd}=1.95\times 10^{-3}$. The disk 
component contributes a relative large fraction (0.16) to the total luminosity.

A fit to J1753.5-0127 with larger viscosities has also been carried out, and 
the relevant results are listed in Table~\ref{t:fits}.  The calculations show that at 
a fixed effective temperature, for larger viscosities the conduction cooling 
to the corona is less efficient and hence the Compton cooling becomes more 
important. Specifically, the inner disk becomes smaller, the condensation rate 
decreases, and disk becomes weaker relative to the corona. Future measurements of 
the disk luminosity fraction and the reflection component will provide further 
constraints on the viscosity.

\section{Discussion and implications}

The evaporation/condensation picture for the accretion geometry of black hole X-ray 
transient systems has been investigated to determine its applicability to the weak 
soft thermal component observed in the low hard states of GX339-4 and J1753.5-0127. 
The model for the mass and energy exchange between the corona and an underlying disk 
developed earlier, based on the thermal conduction of energy, has been extended 
to include the effects of coronal cooling associated with the inverse Compton scattering
process of soft X-ray photons.  The numerical solutions for the disk structure reveal 
that the Compton dominated cooling region is more important for higher soft photon 
fluxes at all radii, with the radial extent $r_{\rm cmp}$ of the region
independent of the mass of the accreting black hole component in the system.  In 
addition, it is found that cool inner disks, contributing a small fraction of the 
total X-ray luminosity ($<20\%$), can exist in the low hard state of black 
hole transient systems GX 339-4 and J1753.5-0127. 

The fits to GX 339-4 and Swift J1753.5-0127 reveal that the theoretical 
disk temperatures are lower than the observationally inferred values, indicating
that additional heating of the disk is required. We suggest that this heating 
is associated with coronal irradiation. 
                                                                                            
Specifically, for a corona in the form of a slab lying above the disk, nearly 
half of the total coronal emission impinges on the optically thick disk and is 
reprocessed as blackbody radiation (Haardt \& Maraschi 1991; Liu et al. 2003). The 
luminosity associated with the irradiation is close to the observed luminosity. 
For GX 339-4, this irradiating luminosity is about 0.03 times the Eddington value.
Its corresponding heating is equivalent to viscous dissipation in the disk with an 
additional accretion rate of 0.03 of the Eddington value, producing an effective 
temperature of about 0.27keV (from  Eq.\ref{T_eff}). Taking into account the viscous 
dissipation from the accretion of condensing matter, the inner disk temperature can 
be similar to the observationally inferred value of 0.3keV. Similarly, for Swift J1753.5-0127,  
$L/L_{\rm edd}= 0.01$, and the disk is heated by irradiation to about 0.2keV, again,
comparable to that observed.
                                                                                           
The inclusion of irradiation does not affect the fits described in \S 4 since the 
disk temperature is fixed to the observed value. Although the effective temperature 
established by viscous dissipation is lower than 0.3keV, the additional heating from 
irradiation can lead to improved comparisons between the theoretically derived 
and observed temperatures.

\subsection{Range of luminosity over which cool inner disks exist}
In the low hard state the luminosity range over which cool inner disks exist depends 
on whether condensation can occur.  To determine the lower limit to the range 
we make use of the fact that the critical condensation radius should fulfill the 
condition that $r_d> 3$.  In the case that conduction cooling dominates in the corona, 
this condition requires 
\begin{equation}\label{lower-L1}
\dot m> 1.63\alpha^{7/2}.
\end{equation} 
For a fixed viscosity parameter, $\alpha=0.2$, a lower limit to the accretion rate for 
an inner disk to exist is $\dot m=0.006$. The corresponding luminosity in this critical 
case is from a pure ADAF luminosity (rather than Bremsstrahlung), which could be as 
small as ${L\over L_{\rm Edd}}=0.001$, depending on the radiative efficiency of the 
ADAF accretion.  At an accretion rate slightly higher than the critical value, the disk 
component, as measured by the disk size or by the disk luminosity fraction for example,
could be very weak and hence Compton cooling is negligible.  We note that this lower 
critical mass accretion rate is very sensitive to $\alpha$.  

For cases in which Compton cooling dominates in the corona, the lower limits 
to the accretion rate for condensation to a disk are greater than that given by 
the conduction model for $0.1< \alpha <0.4$.  In particular, the lower limit to 
the accretion rate is 0.0245 (0.07) and to the luminosity is 0.008 (0.016) for 
$\alpha = 0.2 (0.3)$.

Thus, the lower limit on the mass accretion rate or luminosity leading to the 
formation of an inner disk is set by the conduction model, which is found to 
depend on the viscosity parameter.  For a  $10 M_\odot$ black hole, condensation 
could occur at an accretion rate as low as $\dot m=0.006$ for a standard 
viscosity parameter $\alpha=0.2$.  However,  if observations suggest high-temperature 
soft component (say, $T_ {\rm eff}=0.14$keV), Compton scattering must play a  role 
in cooling of the corona, and the lower limit to the accretion rate is higher,  
corresponding to about 0.01$L_{\rm Edd}$, depending on the viscosity.
  
Hence, our considerations of the two cooling regimes suggest cool inner disks may be 
present for systems with X-ray luminosities as low as 0.001-0.01 $\ledd$. This is 
of particular interest for J1753.5-0127 since it remained in a hard state throughout
its outburst, suggesting that a spectral transition from a hard state to a soft state 
during an outburst is not a necessary condition for the formation of an inner disk 
in the hard state.  Observations of J1753.5-0127 at even lower luminosity levels 
would be especially valuable in constraining the model parameters further.  
Of particular interest, we note that the observations of different classes 
of systems suggest that the viscosity parameter is relatively high ($\alpha \sim 0.1-0.4$, see
King et al. 2007; exceeding more than 0.5 from MHD simulation by Sharma et al. 2006), 
overlapping with the range of values inferred from our modeling.

An upper bound on the luminosity for which an inner disk exists in the hard state
can also be estimated. The luminosity given by the condensation model increases with 
increasing accretion rate and decreasing viscosity.  Thus, the upper bound is 
constrained by the highest accretion rate and lowest viscosity.  For example, at a 
transition accretion rate of 0.02 (0.0675), the upper bound to the luminosity is 
0.003 (0.009) for $\alpha=0.2 (0.3)$.  We point out that at the upper limit of 
$\dot m=0.1$ suggested by Narayan et al. (1995b) for the existence of an ADAF, 
the lowest viscosity is 0.342.  A lower value for $\alpha$ would correspond to 
a transition rate lower than 0.1 (from $\dot m_{\rm tr}\propto \alpha^3$), and 
the system would no longer lie in the low/hard state.  Hence, in this limit, an 
upper bound on the luminosity is 0.02 for this transition accretion rate. 
This is in good agreement with a  hard-to-soft transition luminosity of 1\%-4\% found 
in X-ray binaries (Maccarone 2003).

\subsection{A mass-independent process}
It has been shown in previous investigations of the energy and mass exchange
between the accretion disk and its corona that the results of both the 
evaporation process, i.e., the Eddington scaled evaporation rate ($\dot M_{\rm 
evap}/\dot M_{Edd}$) and its distribution with respect to the scaled distance 
($R/R_{\rm S}$) and the condensation process to a truncated inner cool disk are 
not dependent on the mass of the black hole (Liu et al. 2002; Liu et al. 2006; 
Meyer et al. 2007; see also Eqs.\ref{r_cnd1} and \ref{condensation_cnd}). By 
including the effect of Compton cooling in the present study,  the condensation 
rate appears to depend on the mass of the black hole and the disk effective 
temperature, as shown in Eqs.\ref{r_cnd2} and \ref{condensation-cmp}. If the 
accretion of the condensing matter is the only energy source for establishing 
the thermal properties of the disk,  the effective temperature is no 
longer a free parameter, but is determined by Eq.\ref{T_eff}. Replacing $T_{\rm 
eff,max}$ in Eqs. \ref{r_cnd2}, \ref{A}, and \ref{B} by Eq.\ref{T_eff}, we 
find that the critical radius where condensation takes place, $r_d$,  
and the quantities $A$ and $B$ only depend on the disk viscosity parameter 
and the mass accretion rate as given by 

\begin{equation}\label{r_cnd3}
r_{\rm d}\[1-\({3\over r_d}\)^{1/2}\]^{-4/7}=
311.43\alpha^{-4/3}\dot m^{8/21}\dot m_{\rm cnd}^{4/7},
\end{equation}
\begin{equation}
A=7.174\times 10^{-4}\alpha^{-7/5}\dot m^{7/5}\dot m_{\rm cnd}^{-2/5},
\end{equation}
\begin{equation}\label{B2}
B=25.785\alpha^{-14/15}\dot m^{4/15}\dot m_{\rm cnd}^{2/5}\({r\over 3}\)^{1/2}.
\end{equation}
With above expression for $r_d$, $A$ and $B$  we obtain from Eq.\ref{condensation-cmp}
a non-linear equation describing the  condensation rate $\dot m_{\rm cnd}(r)$, 
which  depends  on $\alpha$ and the Eddington scaled accretion rate $\dot m$, 
but independent of $m$.  For any given accretion rate and viscosity parameter, 
the condensation rate is determined by Eq.\ref{condensation-cmp}.  Likewise, the 
extent of the Compton cooling dominant region, the electron temperature in the 
corona, and the Compton luminosity are determined as
\begin{equation}
r_{\rm cmp}\[{1-\({3\over r_{\rm cmp}}\)^{1/2}}\]^{-2/3}\le 846.71\dot m_{\rm cnd}^{2/3},
\end{equation}
\begin{equation}
T_{\rm ec}=3.52\times 10^8{\rm K}\alpha^{-2/5}\dot m^{2/5}\dot m_{\rm cnd}^{-2/5}r^{1/5}\[1-\({3\over r}\)^{1/2}\]^{-2/5},
\end{equation}
\begin{equation}
{L_{\rm cmp}\over L_{\rm Edd}}=9.88\alpha^{-7/5}\dot m^{7/5}\dot m_{\rm cnd}^{3/5}\int_{r_{\rm d1}/3}^{r_{\rm d2}/3}x^{-23/10}\(1-x^{-1/2}\)^{3/5}dx.
\end{equation}
Thus the condensation process and the thermal state of the corona (i.e., the electron 
temperature) are independent on the mass of the black hole. Taking into account the 
mass-independent features of the previously developed model and its extension in this 
study, we conclude that the simple description of the accretion process within the framework 
of an evaporation/condensation picture can be applied not only to stellar mass black 
holes in X-ray binary systems, but also to supermassive black holes in active galactic 
nuclei. The presence of a cool
disk in the innermost regions surrounding a supermassive black hole in a low state would
directly contribute to the ultraviolet and optical emission and possibly
indirectly via reflected X-ray emission from a spatially extended inner disk. In
addition, it would provide a natural site of cool material where neutral
iron could be present for the production of Fe fluorescent line emission at 6.4 keV
(e.g. Tanaka et al. 1995). A more detailed study of the properties of such a cool disk
would be especially illuminating for interpretations of Fe line profiles based on general
relativistic broadening, though at present it is difficult to observe such a line 
spectrum from low luminosity active galactic nuclei.

\acknowledgments 
This work was supported, in part, by the Theoretical Institute 
for Advanced Research in Astrophysics (TIARA) operated under
Academia Sinica and the National Science Council Excellence 
Projects program in Taiwan administered through grant number NSC 
95-2752-M-007-006-PAE. B.F. Liu acknowledges the support by
the National Natural Science Foundation of China (NSF-10533050)
and the BaiRenJiHua program of the Chinese Academy of Sciences.

\clearpage
\begin{deluxetable}{ccccccc}
\tablecaption{\label{t:examples} The condensation features of disk corona 
accretion around a $10M_\odot$ black hole. For given values of  viscosity 
($\alpha$) and accretion rate ($\dot m$),  the inner disk size ($r_d$), the 
total condensation rate ($\dot m_{\rm cnd}$) and its corresponding maximal 
temperature ($T_{\rm eff,max}$) are listed. The luminosity  ($L/L_{\rm Edd}$) 
from Bremsstrahlung and Compton scattering (if Compton cooling sets in) and 
the dominant cooling mechanism or Compton dominant region  are also shown.}
\tablehead{$\alpha$&$\dot m$&$r_d$&$\dot m_{\rm cnd}$&$T_{\rm eff,max}$(keV)&$L/L_{\rm Edd}$&Cooling mechanism}
\startdata
0.2&0.01&12.6&$4.70\times 10^{-5}$&0.05&$6.52\times 10^{-4}$&Conduction\\
0.3&0.04&11.6&$1.60\times 10^{-4}$&0.07&$2.45\times 10^{-3}$&Conduction\\
0.4&0.08&5.0&$3.32\times 10^{-5}$&0.03&$1.83\times 10^{-3}$&Conduction\\
0.2&0.03\tablenotemark{*}&237&$5.24\times 10^{-3}$&0.17&$1.57\times 10^{-2}$&$3.3<r_{\rm cmp}<18$\\
0.35&0.11&41&$3.00\times 10^{-3}$&0.15&$2.80\times 10^{-2}$&$3.7<r_{\rm cmp}<10.6$\\
0.4&0.16&30.5&0.01&0.20&0.069\tablenotemark{**} &Compton\\
\enddata
\tablenotetext{*}{This is shown as an example for Compton cooling with the standard value of $\alpha$, but the accretion rate exceeds the transition rate.}
\tablenotetext{**}{This luminosity is over-estimated by the increase of optical depth of the corona.} 
\end{deluxetable}

\clearpage

\begin{table}
\begin{center}
\caption{\label{t:fits}The fitting results for GX 339-4 and J1753. For every given 
value of $\alpha$, an accretion rate, $\dot m$, is found, which predicts a 
luminosity consistent with the observational inferred value.  $r_d$, $\dot 
m_{\rm cnd}$, $L/L_{\rm Edd}$, $L_{\rm d}/L_{\rm c}$ and  $T$(keV) are the 
model predictions for the disk size, condensation rate, luminosity, 
ratio of the disk and corona luminosities, and the maximal effective temperature 
caused by accretion.}
\begin{tabular}{ccccccc}
\tableline\tableline
$\alpha$&$\dot m$&$r_d$&$\dot m_{\rm cnd}$&$L/L_{\rm Edd}$&$L_{\rm d}/L_{\rm c}$&$T$(keV)\\
\tableline
\multicolumn{7}{c}{GX 339-4: $L/L_{\rm Edd}=0.03$, $M=10M_\odot$, $T_{\rm eff, max}=0.3$keV}\\
0.2&0.024&3.01--99.5&$6.66\times 10^{-3}$&0.03&21.8\%&0.185\\
0.3&0.037&3.02--66.5&$3.78\times 10^{-3}$&0.03&12.7\%&0.160\\
0.4&0.051&3.03--50.0&$2.54\times 10^{-3}$&0.03&8.4\%&0.145\\
\tableline
\multicolumn{7}{c}{J1753.5-0127: $L/L_{\rm Edd}=0.01, M=10M_\odot$, $T_{\rm eff, max}=0.2$keV}\\
0.2&0.02&3.07--33.4&$1.95\times 10^{-3}$&0.01&18.9\%&0.136\\
0.3&0.034&3.12--22.5&$9.91\times 10^{-4}$&0.01&9.6\%&0.115\\
0.4&0.049&3.2--16.7&$5.73\times 10^{-4}$&0.01&5.7\%&0.100\\
\tableline
\end{tabular}
\end{center}
\end{table}
\clearpage

\end{document}